\documentclass[preprint,authoryear,10pt]{elsarticle}

\usepackage{natbib}
\usepackage{epsfig}
\usepackage{graphicx}
\usepackage{bm}
\usepackage{amsmath,amssymb,amsthm,amsfonts,mathrsfs,textcomp,mathtools}
\usepackage{url}
\usepackage{multirow}
\usepackage{subfig}
\usepackage{aas_macros}

\usepackage{booktabs}
\usepackage{fullpage}
\usepackage{hyperref}

\journal{Celestial Mechanics and Dynamical Astronomy}
\newcommand{\sgn}{\mathop{\mathrm{sgn}}}

\begin{document}

\title{The evolution of the Line of Variations at close encounters: an
  analytic approach}

\author[iaps,ifac]{Giovanni Battista Valsecchi}
\ead{giovanni@iaps.inaf.it}
\author[sds,pi]{Alessio Del Vigna}
\author[cran]{Marta Ceccaroni}

\address[sds]{Space Dynamics Services s.r.l., via Mario Giuntini,
  Navacchio di Cascina, Pisa, Italy}

\address[pi]{Dipartimento di Matematica, Universit\`a di Pisa, Largo
  Bruno Pontecorvo 5, Pisa, Italy}

\address[cran]{Cranfield University, College Road, Cranfield MK43 0AL (UK)}

\address[iaps]{IAPS-INAF, via Fosso del Cavaliere 100, 00133 Roma,
  Italy}

\address[ifac]{IFAC-CNR, via Madonna del Piano 10, 50019 Sesto
  Fiorentino, Italy}

\date{Received xxx; accepted xxx}

\begin{small}
    \begin{abstract}
        We study the post-encounter evolution of fictitious small
        bodies belonging to the so-called Line of Variations (LoV) in
        the framework of the analytic theory of close encounters.  We
        show the consequences of the encounter on the local minimum of
        the distance between the orbit of the planet and that of the
        small body, and get a global picture of the way in which the
        planetocentric velocity vector is affected by the encounter.
        The analytical results are compared with those of numerical
        integrations of the restricted 3-body problem.
    \end{abstract}
\end{small}
\maketitle

\begin{small}{\noindent\bf Keywords}: Close encounter, Perturbation
\end{small}

\section{Introduction}    \label{introduction}

In the framework of their extension of the analytic theory of close encounters originally formulated by
\cite{o76}, \cite{vmgc03} introduced the so-called ``wire approximation'', a simple analytic description of
the Line of Variations (LoV) in which the orbital uncertainty of a small body encountering a planet is
modelled assuming that its orbital parameters $a$, $e$, $i$, $\Omega$ and $\omega$ are constant, and all of
the uncertainty is in the timing of the encounter, i.e., in the mean anomaly $M$.

This rather simplified description differs from other more sophisticated modelisations like those described in
\cite{m99} and \cite{mcstv05}, implemented in the software robots
CLOMON2\footnote{https://newton.spacedys.com/neodys/} and
Sentry\footnote{https://cneos.jpl.nasa.gov/sentry/}, which are in charge of determining whether a given
Near-Earth Asteroid (NEA) can impact the Earth in the coming century.  These software robots take into account
the uncertainties in all the orbital elements, and make accurate propagations in time, including all the known
perturbations, while in the wire approximation the only uncertainty taken into account is that on the timing
of the encounter, which in turn is computed in a very simplified model.  On the other hand, the simple
analytical formulation of the wire approximation allows one to capture some important features of the problem.

The outcome of a planetary fly-by of a planet-crossing small body strongly depends on its coordinates on the
``target plane'', or $b$-plane, of the encounter \citep{k61,gcv88,var18}, i.e. the plane centred on the planet
and perpendicular to the planetocentric velocity ``at infinity'' of the small body.  The uncertainty of the
post-encounter trajectory is a function of the uncertainty in the orbital elements at the time of the
encounter, and in most cases of interest is dominated by the uncertainty in the time of closest approach.  A
suitable choice of the target plane coordinates is such that one coordinate represents the local minimum
distance between the orbit of the small body and that of the planet, and the other is proportional to the
timing of the encounter.  In this way, the uncertainty is mostly along a line parallel to one of the
coordinate axes, and can be modelled by the so-called Line of Variations (LoV). The LoV approach is a crucial
ingredient of the Impact Monitoring software developed at the University of Pisa and at the JPL.

In this paper we show some geometric features of the wire approximation that are deducible from its analytic
formulation and that can be useful to interpret the outcomes of impact monitoring computations coming from
sophisticated, accurate models of the motion of NEAs on Earth approaching orbits.

The paper is organized as follows: in Section~\ref{theory} we briefly recall the analytic theory and how the
LoV is described by the wire approximation; in Section~\ref{postencmoid} we discuss what happens to the LoV
at a close encounter. Afterwards, in Section~\ref{rotation} we give the overall geometric picture of how the
planetocentric velocity vectors of the points belonging to the LoV are rotated, and in Section~
\ref{conclusions} we summarize the conclusions.

\section{Analytic theory of close encounters}  \label{theory}

The analytic theory of close encounters has been developed over the years, starting from \cite{o76}, in a
sequence of papers \citep{gcv88,cvg90,vmgc03,v06,var15a,var15b}, to which we refer the reader.

The basic assumptions are that the small body is massless, and the planet moves on a circular orbit about the
Sun, similarly to what is assumed in the restricted, circular, $3$-dimensional $3$-body problem; however, far
from the planet, the small body is assumed to move on an unperturbed heliocentric Keplerian orbit, not being
subject to the perturbation by the planet. Then, when a close encounter with the planet takes place, the
interaction is modelled as an instantaneous transition from the incoming asymptote of the planetocentric
hyperbola to the outgoing one, taking place when the small body crosses the $b$-plane
\citep{k61,o76,gcv88,cvg90}.

\begin{figure}
\centering
\includegraphics[width=95mm]{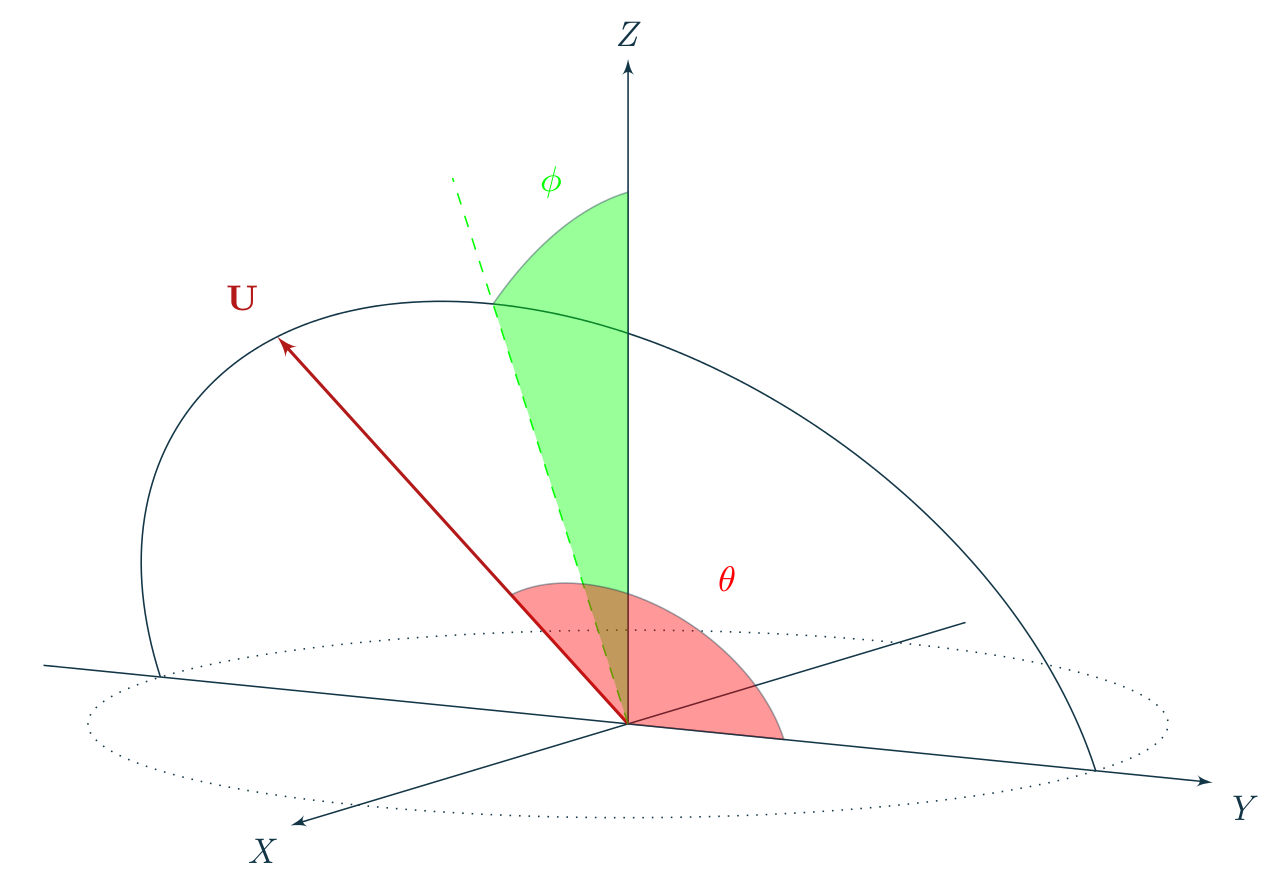}
\caption{The planetocentric velocity vector $\vec{U}$ in the reference frame $X$-$Y$-$Z$; the $X$-axis points
towards the direction opposite to that of the Sun, and the $Y$-axis coincides with the direction of the
heliocentric motion of the planet.  The angle between the $Y$-axis and $\vec{U}$ is $\theta$, and that between
the $Y$-$Z$ plane and the plane containing the $Z$-axis and $\vec{U}$ is $\phi$.}
\label{opik_theta_phi}
\end{figure}

The computation relies on some intermediate variables, the so-called \"Opik variables, constituted by:
\begin{itemize}
\item
the planetocentric velocity vector $\vec{U}$, whose components in a reference frame centred on the planet,
with the $X$-axis pointing away from the Sun, and the $Y$-axis in the direction of the planet motion, are
given by $U_x=U\sin\theta\sin\phi$, $U_y=U\cos\theta$, and $U_z=U\sin\theta\cos\phi$
(see Fig.~\ref{opik_theta_phi});
\item
the two $b$-plane coordinates $\xi$ and $\zeta$;
\item
the time $t_b$ at which the $b$-plane is crossed.
\end{itemize}

As a consequence of an encounter the direction of $\vec{U}$ changes but its modulus $U$ does not; explicit
expressions linking the pre-encounter to the post-encounter orbital parameters, making use of \"Opik
variables, are given in \cite{cvg90}, \cite{vmgc03} and \cite{v06}.

\subsection{The $b$-plane}

As was already mentioned, the $b$-plane of an encounter is the plane containing the planet and perpendicular
to the planetocentric unperturbed velocity $\vec{U}$. The vector from the planet to the point in which
$\vec{U}$ crosses the plane is $\vec{b}$, and the coordinates on the $b$-plane are $\xi$ and $\zeta$.  As
defined in \cite{vmgc03} and \cite{v06}, $\xi=\xi(a,e,i,\omega,\sgn{f_b})$ is the local MOID (Minimum Orbital
Intersection Distance), and $\zeta=\zeta(a,e,i,\Omega,\omega,f_b,\lambda_p)$ is related to the timing of the
encounter. In these expressions, $a$, $e$, $i$, $\Omega$, $\omega$ are the semimajor axis, eccentricity,
inclination, longitude of node and argument of perihelion of the pre-encounter orbit of the small body, and
$f_b$, $\lambda_p$ are the true anomaly of the small body and the longitude of the planet at the crossing of
the $b$-plane.

\subsection{The wire approximation}

In the wire approximation we consider the encounter of a stream of small bodies spaced in mean anomaly (that
is, in true anomaly and therefore in $\zeta$), all on the same orbit, with given local MOID $\xi_0$.  Then, as
previously mentioned, $U$ does not change as a result of the close encounter and $t_b$ does not concern us
here.

The encounter changes the angles $\theta$ and $\phi$ into $\theta'$ and $\phi'$.  Moreover, we can consider
that for each pair $\theta',\phi'$ we can define an ``outgoing'' $b$-plane, normal to post-encounter velocity
vector $\vec{U}'$\footnote{The components of $\vec{U}'$ in the $X$-$Y$-$Z$ reference frame are
($U\sin\theta'\sin\phi',~U\cos\theta',~U\sin\theta'\cos\phi'$).}, that is crossed by the small body at
coordinates $\xi'$ and $\zeta'$. We call this plane the ``post-encounter $b$-plane''.

\cite{vmgc03} give the relevant equations to compute the post-encounter quantities of the small bodies along
the wire:
\begin{eqnarray}
\cos\theta' & = & \frac{(\xi_0^2+\zeta^2-c^2)\cos\theta+2c\zeta\sin\theta}
{\xi_0^2+\zeta^2+c^2} \label{ctp} \\
\sin\theta' & = & \frac{\sqrt{[(\xi_0^2+\zeta^2-c^2)\sin\theta-2c\zeta\cos\theta]^2+4c^2\xi_0^2}}
{\xi_0^2+\zeta^2+c^2} \label{stp} \\
\sin\phi' & = & \frac{[(\xi_0^2+\zeta^2-c^2)\sin\theta-2c\zeta\cos\theta]\sin\phi-2c\xi_0\cos\phi}
{(\xi_0^2+\zeta^2+c^2)\sin\theta'} \label{spp} \\
\cos\phi' & = & \frac{[(\xi_0^2+\zeta^2-c^2)\sin\theta-2c\zeta\cos\theta]\cos\phi+2c\xi_0\sin\phi}
{(\xi_0^2+\zeta^2+c^2)\sin\theta'} \label{cpp} \\
\xi' & = & \frac{\xi_0\sin\theta}{\sin\theta'} \\
\zeta' & = & \frac{(\xi_0^2+\zeta^2-c^2)\zeta\sin\theta-2(\xi_0^2+\zeta^2)c\cos\theta}
{(\xi_0^2+\zeta^2+c^2)\sin\theta'},
\end{eqnarray}
with $c$ given by:
\begin{equation}
c=\frac{m}{U^2},
\end{equation}
where $m$ is the mass of the planet in units of that of the Sun.  Particularly noteworthy is the expression
for $\xi'$, which gives the post-encounter local MOID; we discuss its implications in Sect.~\ref{postencmoid}.

\section{Post-encounter local MOID along the wire} \label{postencmoid}

On the post-encounter $b$-plane the size of the post-encounter impact parameter $b'$must be the same as that
of the pre-encounter one $b$, due to the conservation of the planetocentric orbital angular momentum; thus,
the post-encounter local MOID $\xi'$ is bounded:
\begin{displaymath}
0\leq\xi'\leq{b}=\sqrt{\xi^2+\zeta^2}.
\end{displaymath}
Moreover, since $\theta$ and $\theta'$ take values between $0^\circ$ and $180^\circ$ \citep{cvg90}, $\xi$ and
$\xi'$ have the same sign (we remind the reader that $\xi$ and $\xi'$ are coordinates on different planes).

Let us now discuss the variation in size of the local MOID due to the encounter, for a wire that has
$\xi=\xi_0$.  Equations~(\ref{ctp}) and (\ref{stp}) show that, the larger the value of $|\zeta|$, the closer
$\theta'$ will be to $\theta$, and thus the closer $\xi'$ will be to $\xi_0$.

For smaller values of $|\zeta|$, there must be a minimum and a maximum value of $\sin\theta'$, that
correspond respectively to the maximum and minimum values of $\xi'$. To find them, let us consider the
derivative of $\sin\theta'$ with respect to $\zeta$:
\begin{equation}
\frac{\partial\sin\theta'}{\partial\zeta}
=\frac{\partial\sin\theta'}{\partial\cos\theta'}\frac{\partial\cos\theta'}{\partial\zeta}
=-\frac{\cos\theta'}{\sin\theta'}\frac{\partial\cos\theta'}{\partial\zeta}.
\end{equation}
The zeroes of $\partial\sin\theta'/\partial\zeta$ include those of $\partial\cos\theta'/\partial\zeta$, as
well as the values of $\zeta$ such that $\cos\theta'=0$.  As regards the zeroes of
$\partial\cos\theta'/\partial\zeta$, these can be found by zeroing the numerator of this derivative, since
its denominator cannot be negative:
\begin{eqnarray*}
\frac{\partial\cos\theta'}{\partial\zeta} & = & \frac{2c[2c\zeta\cos\theta+(\xi_0^2-\zeta^2+c^2)\sin\theta]}
{(\xi_0^2+\zeta^2+c^2)^2} \\
\zeta_\pm & = & \frac{c\cos\theta\pm\sqrt{c^2+\xi_0^2\sin^2\theta}}{\sin\theta}.
\end{eqnarray*}
Concerning the values of $\zeta$ such that $\cos\theta'=0$, they are given by the intersections of the
straight line $\xi=\xi_0$ with the circle having coordinates of the centre $(0,D)$ and radius $|R|$, given by
\citep{vmgc00,vmgc03}:
\begin{eqnarray*}
D & = & -\frac{c\sin\theta}{\cos\theta} \\
R & = & -\frac{c}{\cos\theta}.
\end{eqnarray*}
The equation of the circle is:
\begin{displaymath}
\xi^2+\zeta^2-2D\zeta+D^2=R^2.
\end{displaymath}
Its intersections with the straight line $\xi=\xi_0$ are the roots of the equation:
\begin{displaymath}
0=\zeta^2+\frac{2c\zeta\sin\theta}{\cos\theta}+\xi_0^2-c^2,
\end{displaymath}
that are:
\begin{displaymath}
\zeta_{1,2}=\frac{-c\sin\theta\pm\sqrt{c^2-\xi_0^2\cos^2\theta}}{\cos\theta}.
\end{displaymath}
Summarizing, the zeroes of $\partial\sin\theta'/\partial\zeta$ are:
\begin{eqnarray}
\zeta_\pm & = & \frac{c\cos\theta\pm\sqrt{c^2+\xi_0^2\sin^2\theta}}{\sin\theta} \\
\zeta_{1,2} & = & \frac{-c\sin\theta\pm\sqrt{c^2-\xi_0^2\cos^2\theta}}{\cos\theta}.
\end{eqnarray}
Note that for
\begin{equation}
|\xi_0|>|R|=\frac{c}{|\cos\theta|}
\end{equation}
there is no intersection of the circle corresponding to $\cos\theta'=0$ with the straight line $\xi=\xi_0$,
so that there are no real values for the roots $\zeta_{1,2}$.

The values of $\xi'$ corresponding to $\zeta_\pm$ are:
\begin{equation}
\xi'_\pm=\frac{\xi_0}{|\xi_0|}\cdot\frac{\sqrt{c^2+\xi_0^2\sin^2\theta}\pm{c}\cos\theta}{\sin\theta},
\end{equation}
while those of $\xi'$ corresponding to $\zeta_{1,2}$, when present, are:
\begin{equation}
\xi'_{1,2}=\xi_0\sin\theta.
\end{equation}

To see how the above expressions work in practice, let us apply the analytical theory to the encounter of
2012~TC$_4$ with the Earth that has taken place on 12 October 2017.  The geocentric velocity of 2012~TC$_4$ is
$U=0.235$, relatively low for a NEA, making this case rather challenging for the analytic theory, that works
best for encounters in which the planetocentric velocity is high.

\begin{figure}
\centering
\includegraphics[width=95mm]{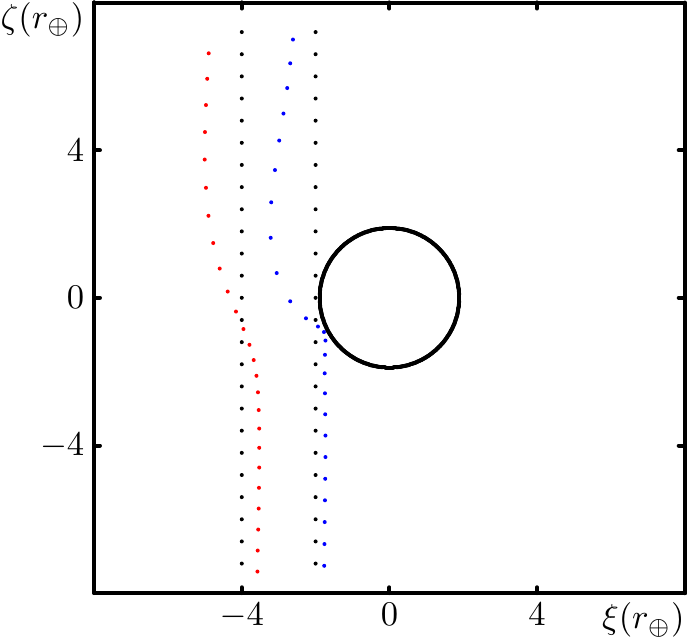}
\caption{The Earth encounter of 2012~TC$_4$ on 12/10/2017; the plot shows the deformation of the LoV for
$\xi_0=-4$ and $\xi_0=-2$ Earth radii.  The black circle represents the Earth cross-section; the black dots
are the points belonging to the two LoVs.  The red dots show the corresponding points in the post-encounter
$b$-plane for $\xi_0=-4$; the blue dots do the same for $\xi_0=-2$.}
\label{fig_lov1}
\end{figure}

Figure~\ref{fig_lov1} shows the $b$-plane relative to this encounter; in it, the black circle centred in the
origin is the gravitational cross-section of the Earth, and the unit adopted for the axes is the physical
radius of our planet $r_\oplus$.  In the case of 2012~TC$_4$, the effective radius of the Earth, due to
gravitational focussing, is $1.89r_\oplus$.  The black dots represent LoV points for two different values of
$\xi$, namely $\xi_0=-4$ and $\xi_0=-2$ Earth radii; the red dots show the post-encounter values $\xi',\zeta'$
corresponding to each pair $\xi_0, \zeta$, for $\xi_0=-4$ Earth radii, and the blue dots do the same for
$\xi_0=-2$ Earth radii.  As already said, to each pair $\xi_0, \zeta$ corresponds a post-encounter pair
$\xi', \zeta'$ defined on a different post-encounter $b$-plane; here, however, we plot them on the
pre-encounter $b$-plane in order to show how the LoV is deformed as a consequence of the close encounter. It
is noteworthy how the variation of the local MOID can be, at least in this case, comparable to the radius of
the Earth.

\begin{figure}
\centering
\includegraphics[width=95mm]{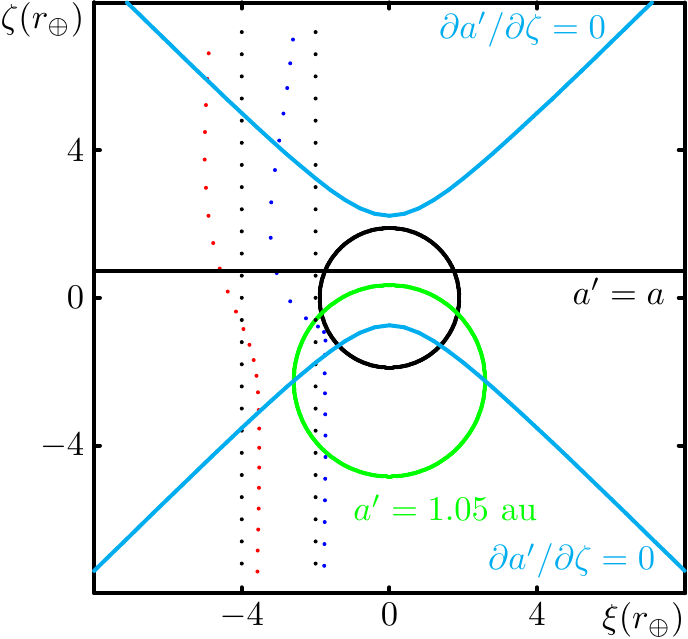}
\caption{Same as Fig.~\ref{fig_lov1}, highlighting relevant $b$-plane loci \citep{var18}.  The cyan hyperbola
corresponds to $\partial\cos\theta'/\partial\zeta=0$; the black straight line is the condition for
$\theta'=\theta$ \citep{vmgc00}; the green circle is the condition $\cos\theta'=0$.}
\label{lovs_loci}
\end{figure}

Figure~\ref{lovs_loci} is similar to Fig.~\ref{fig_lov1}, but also shows the relevant $b$-plane loci
\citep{var18} whose intersections with the LoVs give origin to specific values of $\xi'$.  These loci are:
\begin{itemize}
\item
the condition $\partial\cos\theta'/\partial\zeta=0$, shown by the cyan hyperbola;
\item
the condition $\theta'=\theta$ \citep{vmgc00}, implying $a'=a$ and $\xi'=\xi$, shown by the black horizontal
straight line;
\item
the condition $\cos\theta'=0$ (in this particular case giving   $a'=1.05$~au), corresponding to
$\xi'=\xi\sin\theta$, shown by the green circle.
\end{itemize}
Let us examine the LoV with $\xi_0=-4$ Earth radii, going from positive $\zeta$ values towards negative ones.
For large positive values of $\zeta$, as already said, $\theta'$ tends to $\theta$, so the variation of $\xi$
is small.

Going towards $\zeta=0$, the LoV crosses the hyperbola for $\zeta=\zeta_+$: this corresponds to the maximum of
$\cos\theta'$, i.e. to the minimum of $\sin\theta'$, and thus to the maximum of $|\xi'|$.  The values of
$\theta'$, $\phi'$ and $\xi'$ are given by:
\begin{eqnarray}
\cos\theta'_+ & = & \frac{\sqrt{c^2+\xi_0^2\sin^2\theta}\cos\theta+c}
{\sqrt{c^2+\xi_0^2\sin^2\theta}+c\cos\theta} \label{ct+} \\
\sin\theta'_+ & = & \frac{|\xi_0|\sin^2\theta}{\sqrt{c^2+\xi_0^2\sin^2\theta}+c\cos\theta} \\
\sin\phi'_+ & = & \frac{\xi_0}{|\xi_0|}\cdot\frac{\xi_0\sin\theta\sin\phi-c\cos\phi}
{\sqrt{c^2+\xi_0^2\sin^2\theta}} \label{sp+} \\
\cos\phi'_+ & = & \frac{\xi_0}{|\xi_0|}\cdot\frac{\xi_0\sin\theta\cos\phi+c\sin\phi}
{\sqrt{c^2+\xi_0^2\sin^2\theta}} \label{cp+} \\
\xi'_+ & = & \frac{\xi_0}{|\xi_0|}\cdot\frac{\sqrt{c^2+\xi_0^2\sin^2\theta}+c\cos\theta}{\sin\theta}.
\end{eqnarray}
The next locus encountered by the $\xi_0=-4$ Earth radii LoV is the horizontal straight line corresponding to
$\cos\theta'=\cos\theta$. In this case, the local MOID is unchanged, $\xi'=\xi_0$.

Finally, the LoV encounters the other branch of the hyperbola, in $\zeta_-$; here, the values of $\theta'$,
$\phi'$ and $\xi'$ are given by:
\begin{eqnarray}
\cos\theta'_- & = & \frac{\sqrt{c^2+\xi_0^2\sin^2\theta}\cos\theta-c}
{\sqrt{c^2+\xi_0^2\sin^2\theta}-c\cos\theta} \label{ct-} \\
\sin\theta'_- & = & \frac{|\xi_0|\sin^2\theta}{\sqrt{c^2+\xi_0^2\sin^2\theta}-c\cos\theta} \\
\sin\phi'_- & = & \frac{\xi_0}{|\xi_0|}\cdot\frac{\xi_0\sin\theta\sin\phi-c\cos\phi}
{\sqrt{c^2+\xi_0^2\sin^2\theta}} \label{sp-} \\
\cos\phi'_- & = & \frac{\xi_0}{|\xi_0|}\cdot\frac{\xi_0\sin\theta\cos\phi+c\sin\phi}
{\sqrt{c^2+\xi_0^2\sin^2\theta}} \label{cp-} \\
\xi'_- & = & \frac{\xi_0}{|\xi_0|}\cdot\frac{\sqrt{c^2+\xi_0^2\sin^2\theta}-c\cos\theta}{\sin\theta}.
\end{eqnarray}
A noteworthy feature is that $\phi'_+=\phi'_-$, implying that the extrema of the post-encounter values of
$\theta$ lie on the same meridian.  The difference $\Delta_{max}\theta'$ between them is:
\begin{equation}
\cos(\Delta_{max}\theta')=\frac{\xi_0^2-c^2}{\xi_0^2+c^2}; \label{delta_theta_max}
\end{equation}
we will return on this issue in Sect.~\ref{rotation}.

Coming now to the $\xi_0=2$ Earth radii LoV, the crossings of the hyperbola and of the straight line are as
before. he  However, this LoV crosses also the green circle corresponding to $\cos\theta'=0$; actually, one of
these crossings happens also to take place at the border of the cross-section of the Earth.  Anyway, at these
two crossings we have:
\begin{eqnarray*}
\cos\theta'_{1,2} & = & 0 \\
\sin\theta'_{1,2} & = & 1 \\
\xi'_{1,2} & = & \xi_0\sin\theta.
\end{eqnarray*}

\subsection{Numerical check}

To test the validity of the theoretical predictions about the variation of the local MOID along the LoV,
we proceeded as in Section 4 of \cite{var18}.  That is, we integrated the equations of the restricted,
circular, $3$-dimensional $3$-body problem using the RA15 integrator \citep{e85} with initial conditions
corresponding to the 12 October 2017 encounter of 2012~TC$_4$.

\begin{figure}
\centering
\includegraphics[width=95mm]{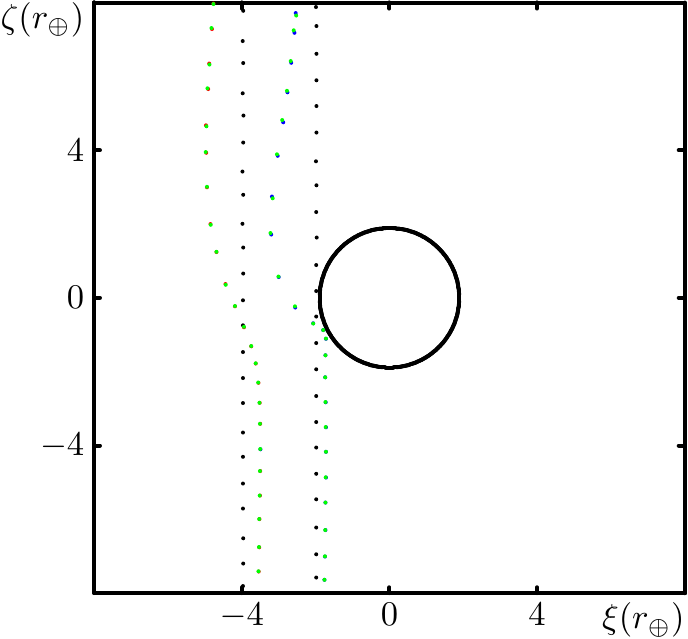}
\caption{Same as Fig.~\ref{fig_lov1}, with the red and blue dots showing the points in the post-encounter
$b$-plane.  The green points, nearly exactly superimposed on the red blue dots, come from the numerical
integrations in the restricted, circular, 3-dimensional 3-body problem, as described in the text; the theory
and the integrations appear to be in very good agreement.}
\label{fig_lov2}
\end{figure}

By trial-and-error we found the pre-encounter values of $\omega$ corresponding to $\xi_0=-4$ and $\xi_0=-2$
Earth radii, and then we integrated sets of initial conditions equally spaced in $\zeta$, thus reproducing the
two LoVs of interest.  Then, we determined the post-encounter values $\xi', \zeta'$ and plotted them in
Fig.~\ref{fig_lov2}, that has to be compared with Fig.~\ref{fig_lov1}.

The theoretical behaviour of the LoV is very well confirmed by the numerical integrations.

\section{Rotation of $\vec{U}$ into $\vec{U}'$ along the wire} \label{rotation}

The conservation of $U$ implies that the pre-encounter and post-encounter velocity vectors $\vec{U}$ and
$\vec{U}'$ span a sphere in $X$-$Y$-$Z$ space, the $U\!$-sphere, of radius $U$ centred in the origin and on
which the angles $\theta$ and $\phi$ define a system of parallels and meridians: on the $U\!$-sphere $\theta$
is the colatitude measured from the $Y$-axis (the direction of motion of the planet), and $\phi$ is the
longitude, counted from the $Z$-$Y$ plane.

Let us examine the path followed by the tip of $\vec{U}'$ on the $U\!$-sphere for a given value of
$\xi_0$.  The angle $\gamma$ between the vectors $\vec{U}$ and $\vec{U}'$ is given by:
\begin{equation}
\tan\frac{\gamma}{2}=\frac{c}{\sqrt{\xi_0^2+\zeta^2}};
\end{equation}
this implies that, for $\zeta\rightarrow\pm\infty$, $\gamma\rightarrow0$.  On the other hand, $\gamma_{max}$,
the maximum value of $\gamma$, is reached for $\zeta=0$, and is given by:
\begin{equation}
\cos\gamma_{max}=\frac{\xi_0^2-c^2}{\xi_0^2+c^2}.  \label{gamma_max}
\end{equation}
A comparison of (\ref{delta_theta_max}) and (\ref{gamma_max}) shows that $\Delta_{max}\theta'$ and
$\gamma_{max}$ are equal; this suggests the possibility that the path followed by the tip of $\vec{U}'$ on the
$U\!$-sphere for a given value of $\xi_0$ might be a circle.

To check whether this is true, let us consider meridian $\phi_P=\phi'_+=\phi'_-$.  On it lie the points $P_+$,
of coordinates $(X_+, Y_+, Z_+)$, and $P_-$, of coordinates $(X_-, Y_-, Z_-)$, where:
\begin{displaymath}
X_+=U\sin\theta'_+\sin\phi'_+ \qquad Y_+=U\cos\theta'_+ \qquad Z_+=U\sin\theta'_+\cos\phi'_+,
\end{displaymath}
and
\begin{displaymath}
X_-=U\sin\theta'_-\sin\phi'_- \qquad Y_-=U\cos\theta'_- \qquad Z_-=U\sin\theta'_-\cos\phi'_-.
\end{displaymath}
These two points correspond to the tips of the post-encounter velocity vector for, respectively,
$\zeta=\zeta_+$ and $\zeta=\zeta_-$; halfway between them, on the same meridian, we now consider point $P$,
whose colatitude $\theta_P$ is halfway between $\theta'_+$ and $\theta'_-$, so that:
\begin{equation}
\theta_P=\frac{\theta'_++\theta'_-}{2}.
\end{equation}
From Eqs.~(\ref{ct+}), (\ref{sp+}), (\ref{cp+}) and (\ref{ct-}) we can compute $\theta_P, \phi_P$ as functions
$c, \theta, \phi, \xi_0$:
\begin{eqnarray}
\cos\theta_P & = & \frac{|\xi_0|\cos\theta}{\sqrt{\xi_0^2+c^2}} \\
\sin\theta_P & = & \frac{\sqrt{\xi_0^2\sin^2\theta+c^2}}{\sqrt{\xi_0^2+c^2}} \\
\sin\phi_P & = & \frac{\xi_0}{|\xi_0|}\cdot\frac{\xi_0\sin\theta\sin\phi-c\cos\phi}
{\sqrt{\xi_0^2\sin^2\theta+c^2}} \\
\cos\phi_P & = & \frac{\xi_0}{|\xi_0|}\cdot\frac{\xi_0\sin\theta\cos\phi+c\sin\phi}
{\sqrt{\xi_0^2\sin^2\theta+c^2}},
\end{eqnarray}
In the $X$-$Y$-$Z$ frame the coordinates of $P$ are:
\begin{eqnarray}
P_X & = & U\sin\theta_P\sin\phi_P
=\frac{\xi_0}{|\xi_0|}\cdot\frac{U(\xi_0\sin\theta\sin\phi-c\cos\phi)}{\sqrt{\xi_0^2+c^2}} \\
P_Y & = & U\cos\theta_P=\frac{U|\xi_0|\cos\theta}{\sqrt{\xi_0^2+c^2}} \\
P_Z & = & U\sin\theta_P\cos\phi_P
=\frac{\xi_0}{|\xi_0|}\cdot\frac{U(\xi_0\sin\theta\cos\phi+c\sin\phi)}{\sqrt{\xi_0^2+c^2}}.
\end{eqnarray}
On the other hand, the post-encounter values $\theta', \phi'$ for a generic initial condition ``on the
wire'', of coordinates $\xi_0, \zeta$, can be computed using Eqs.~(\ref{ctp})-(\ref{cpp}). The corresponding
point on the $U\!$-sphere will have coordinates:
\begin{displaymath}
X=U\sin\theta'\sin\phi' \qquad Y=U\cos\theta' \qquad Z=U\sin\theta'\cos\phi'.
\end{displaymath}

From Eqs.\ (\ref{ctp})-(\ref{cpp}), we rewrite the above expressions as follows:
\begin{eqnarray}
X & = & \frac{U\{[(\xi_0^2+\zeta^2-c^2)\sin\theta-2c\zeta\cos\theta]\sin\phi-2c\xi_0\cos\phi\}}
{\xi_0^2+\zeta^2+c^2} \\
Y & = & \frac{U[(\xi_0^2+\zeta^2-c^2)\cos\theta+2c\zeta\sin\theta]}{\xi_0^2+\zeta^2+c^2} \\
Z & = & \frac{U\{[(\xi_0^2+\zeta^2-c^2)\sin\theta-2c\zeta\cos\theta]\cos\phi+2c\xi_0\sin\phi\}}
{\xi_0^2+\zeta^2+c^2}.
\end{eqnarray}
The square of distance $D_P$ from $P$ to a generic point on the $U\!$-sphere corresponding to an initial
condition ``on the wire'' is:
\begin{equation}
D_P^2=(X-P_X)^2+(Y-P_Y)^2+(Z-P_Z)^2;
\end{equation}
substituting the expressions as functions of $U, \theta, \phi, \xi_0, \zeta$, one obtains:
\begin{equation}
D_P^2=\frac{2U^2\left(\sqrt{\xi_0^2+c^2}-\xi_0\right)}{\sqrt{\xi_0^2+c^2}}.
\end{equation}
Thus the post-encounter values of $\theta'$ and $\phi'$ accessible to a small body encountering the planet
``on the wire'' define the circle resulting from the intersection of the cone of aperture $\gamma_{max}$,
centred in the centre of the $U$-sphere, and the sphere itself.  The pole of the spherical cap delimited by
the circle is point $P$.

For $\xi_0^2\gg{c}^2$, i.e. when the MOID is relatively large, and very close encounters are not possibe,
$\theta_P, \phi_P$ become:
\begin{eqnarray*}
\cos\theta_P & \approx & \cos\theta \\
\sin\theta_P & \approx & \sin\theta \\
\sin\phi_P & \approx & \sin\phi-\frac{c\cos\phi}{\xi_0\sin\theta} \\
\cos\phi_P & \approx & \cos\phi+\frac{c\sin\phi}{\xi_0\sin\theta},
\end{eqnarray*}
that is, $P$ tends towards the tip of $\vec{U}$.  On the other hand, for $\xi_0^2\ll{c}^2$, $\theta_P, \phi_P$
become:
\begin{eqnarray*}
\cos\theta_P & \approx & \frac{|\xi_0|\cos\theta}{c} \\
\sin\theta_P & \approx & 1 \\
\sin\phi_P & = & \frac{|\xi_0|\sin\theta\sin\phi}{c}-\frac{\xi_0\cos\phi}{|\xi_0|} \\
\cos\phi_P & = & \frac{|\xi_0|\sin\theta\cos\phi}{c}+\frac{\xi_0\sin\phi}{|\xi_0|}.
\end{eqnarray*}
Moreover, it is clear that, the smaller becomes $\xi_0$ relative to $c$, the larger becomes the circle spanned
by the initial conditions ``on the wire'' until, for $\xi_0=0$, it becomes the great circle corresponding to
the $\phi$-meridian.  For $\xi_0\neq0$ the circle is tangent to the $\phi$-meridian in the point of spherical
coordinates $\theta, \phi$.

For completeness, we now give expressions for the coordinates of the centre and for the radius $R_C$ of the
circle.  The centre $C$ lies on the axis of the cone, at a distance $U\sin(\gamma_{max}/2)$ from the origin of
axes, while the radius $R_C$ is equal to $U\cos(\gamma_{max}/2)$.  Since:
\begin{displaymath}
\sin\frac{\gamma_{max}}{2}=\frac{c}{\sqrt{\xi_0^2+c^2}} \qquad
\cos\frac{\gamma_{max}}{2}=\frac{|\xi_0|}{\sqrt{\xi_0^2+c^2}},
\end{displaymath}
the coordinates of the $C$ are:
\begin{eqnarray}
C_X & = & P_X\cos\frac{\gamma_{max}}{2}=\frac{U\xi_0(\xi_0\sin\theta\sin\phi-c\cos\phi)}{\xi_0^2+c^2} \\
C_Y & = & P_Y\cos\frac{\gamma_{max}}{2}
=\frac{U\xi_0^2\cos\theta}{\xi_0^2+c^2} \\
C_Z & = & P_Z\cos\frac{\gamma_{max}}{2}=\frac{U\xi_0(\xi_0\sin\theta\cos\phi+c\sin\phi)}{\xi_0^2+c^2},
\end{eqnarray}
and the radius $R_C$ is:
\begin{equation}
R_C=U\sin\frac{\gamma_{max}}{2}=\frac{Uc}{\sqrt{\xi_0^2+c^2}}.
\end{equation}

\begin{figure}[h]
\centering
\includegraphics[width=110mm]{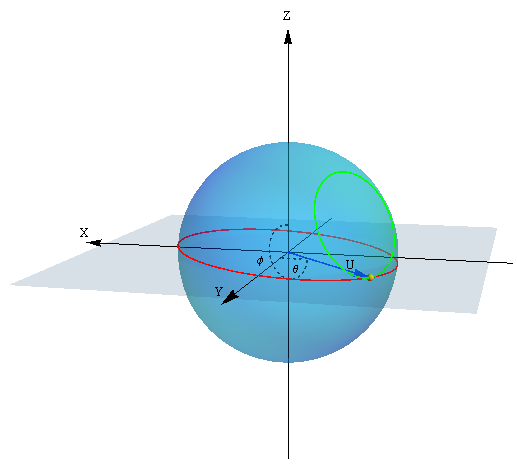}
\caption{The $U$-sphere relative to the 2017 encounter of 2012~TC$_4$ with the Earth.  The red circle is the
intersection of the $U$-sphere with the ecliptic.  The geocentric velocity vector $\vec{U}$, as well as the
angles $\theta$ and $\phi$, are indicated.  The green circle shows the possible directions in which the
post-encounter velocity vector $\vec{U}'$ can be deflected.  The maximum deflection angle $\gamma_{max}$ in
this case is $56^\circ\!\!.8$.}
\label{57}
\end{figure}

As an application of the above considerations, let us consider the already mentioned recent encounter of
2012~TC$_4$ with the Earth.  The relevant quantities in this case are:
\begin{eqnarray*}
U & = & 0.235 \\
\theta & = & 60^\circ\!\!.2 \\
\phi & = & 265^\circ\!\!.3 \\
\frac{c}{r_\oplus} & = & 1.29 \\
\frac{\xi_0}{r_\oplus} & = & -2.38 \\
\cos\gamma_{max} & = & 56^\circ\!\!.8,
\end{eqnarray*}
where $r_\oplus$ is the radius of the Earth.

Figure~\ref{57} helps to visualize the situation.  It shows $\vec{U}$ in the $X$-$Y$-$Z$ frame, as well as the
$U$-sphere, spanned by $\vec{U}'$ for all possible values of $\xi, \zeta$; the red circle is the intersection
of the $U$-sphere with the $X$-$Y$ plane, and the angles $\theta, \phi$ are indicated.  The green circle is
spanned by $\vec{U}'$ for all possible values of $\zeta$ and $\xi=-2.38$ Earth radii.

\begin{figure}[h]
\centering
\includegraphics[width=110mm]{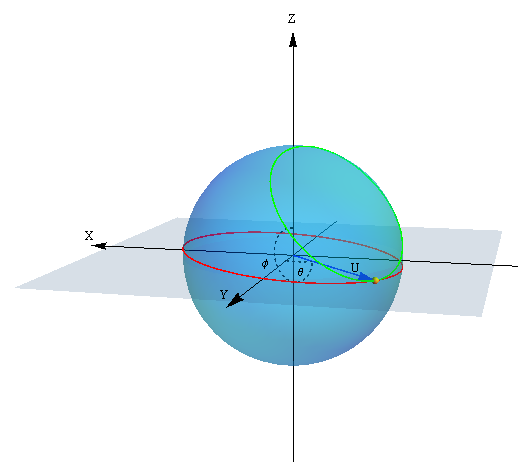}
\caption{Same as Fig.~\ref{57}, for $\xi_0=-c$, so that $\gamma_{max}=90^\circ$.}
\label{90}
\end{figure}

Figure~\ref{90} shows the situation for a value of $\xi_0=-c$; this value is still negative, but closer to
$0$, and for it $\gamma_{max}$ amounts to $90^\circ$.  Note that we are showing the behaviour of $\vec{U}'$
also for deflections that would imply a perigee of the real asteroid smaller than the radius of the Earth, in
order to give the overall view of the geometry involved.  Obviously, in a realistic computation, parts of the
green circle would be forbidden, due to the impact.

\begin{figure}[h]
\centering
\includegraphics[width=110mm]{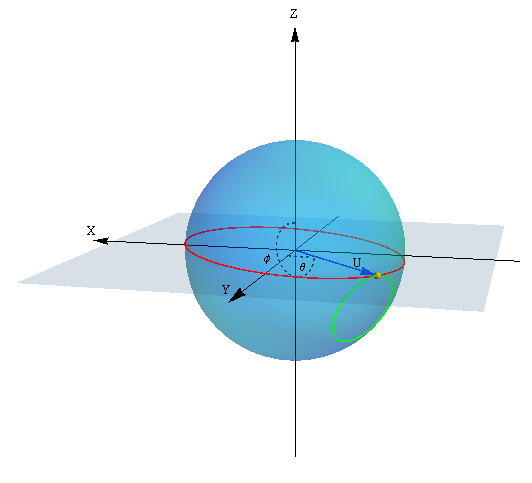}
\caption{Same as Fig.~\ref{57}, for $\xi_0=3.65\cdot{r}_\oplus$, resulting in $\gamma_{max}=38^\circ\!\!.9$.}
\label{321}
\end{figure}

Finally, Fig.~\ref{321} shows what happens when $\xi_0$ changes sign.  In the case shown, $\xi_0=3.65$
terrestrial radii, so that $\gamma_{max}=38^\circ\!\!.9$.  As already said, for $\xi_0=0$ the green circle
becomes a great circle; afterwards, the green circle starts to shrink on the other side, as $ \xi_0$ starts to
increase after having passed through $0$.

\section{Conclusions} \label{conclusions}

We discussed how a close encounter, in which the local MOID is well determined and the timing is somewhat
uncertain, can be modelled with the wire approximation, in which the LoV on the $b$-plane is described by
$\xi=\xi_0$ and $\zeta$ takes any value within the uncertainty range.

Explicit expressions can then be given that describe the behaviour of the LoV after the encounter.  In
particular, these expressions allow one to describe the variation of the local MOID, that in some cases can
be of the order of the radius of the Earth, and thus have consequences for the possibility of subsequent
impacts.

Numerical integrations in the restricted, circular $3$-dimensional $3$-body problem confirm that the
theoretical results on the variation of the local MOID are satisfactorily accurate.

Moreover, the theory allows us to give the overall geometrical description of how the planetocentric velocity
vector is deflected at the encounter, as a function of the MOID of the orbits described by the LoV in the wire
approximation.  In fact, for a LoV of given $\xi_0$, the post-encounter values of $\theta'$ and $\phi'$ lead
to a circle resulting from the intersection of the cone of aperture $\gamma_{max}=\gamma_{max}(\xi_0, c)$,
centred in the centre of the sphere spanned by $\vec{U}'$, and the sphere itself.

Comparisons of these results with those that can be obtained in realistic situations, for real asteroids
possibly impacting the Earth, will be the subject of future work.

\section*{Acknowledgements}
We are grateful to D. Farnocchia for his very useful comments.

Conflict of Interest: The authors declare that they have no conflict of interest.
\bibliographystyle{elsarticle-harv}
\bibliography{LoV}

\end{document}